\documentclass[a4paper,11pt]{article}
\usepackage{jcappub, bm, color} 
\usepackage{amssymb,amsfonts,slashed,amsthm,amsmath,graphicx, soul, empheq}
\usepackage[caption=false]{subfig}
\begin{document}

\newcommand{\vev}[1]{ \left\langle {#1} \right\rangle }
\newcommand{\bra}[1]{ \langle {#1} | }
\newcommand{\ket}[1]{ | {#1} \rangle }
\newcommand{\eV}{ \ {\rm eV} }
\newcommand{\KeV}{ \ {\rm keV} }
\newcommand{\MeV}{\  {\rm MeV} }
\newcommand{\GeV}{\  {\rm GeV} }
\newcommand{\TeV}{\  {\rm TeV} }
\newcommand{\1}{\mbox{1}\hspace{-0.25em}\mbox{l}}
\newcommand{\Red}[1]{{\color{red} {#1}}}

\newcommand{\lmk}{\left(}  
\newcommand{\rmk}{\right)}
\newcommand{\lkk}{\left[}  
\newcommand{\rkk}{\right]}
\newcommand{\lhk}{\left \{ }  
\newcommand{\rhk}{\right \} }
\newcommand{\del}{\partial}  
\newcommand{\la}{\left\langle} 
\newcommand{\ra}{\right\rangle}
\newcommand{\half}{\frac{1}{2}}

\newcommand{\bea}{\begin{array}}
\newcommand{\eea}{\end{array}}
\newcommand{\beq}{\begin{eqnarray}}
\newcommand{\eeq}{\end{eqnarray}}
\newcommand{\eq}[1]{Eq.~(\ref{#1})}

\newcommand{\dd}{\mathrm{d}}
\newcommand{\Mpl}{M_{\rm Pl}}
\newcommand{\mg}{m_{3/2}}
\newcommand{\abs}[1]{\left\vert {#1} \right\vert}
\newcommand{\mphi}{m_{\phi}}
\newcommand{\Hz}{\ {\rm Hz}}
\newcommand{\for}{\quad \text{for }}
\newcommand{\Min}{\text{Min}}
\newcommand{\Max}{\text{Max}}
\newcommand{\Kahler}{K\"{a}hler }
\newcommand{\cphi}{\varphi}
\newcommand{\Tr}{\text{Tr}}
\newcommand{\diag}{{\rm diag}}

\newcommand{\SUf}{SU(3)_{\rm f}}
\newcommand{\Upq}{U(1)_{\rm PQ}}
\newcommand{\Zpq}{Z^{\rm PQ}_3}
\newcommand{\Cpq}{C_{\rm PQ}}
\newcommand{\ubar}{u^c}
\newcommand{\dbar}{d^c}
\newcommand{\ebar}{e^c}
\newcommand{\nubar}{\nu^c}
\newcommand{\Ndw}{N_{\rm DW}}
\newcommand{\Fpq}{F_{\rm PQ}}
\newcommand{\fpq}{v_{\rm PQ}}
\newcommand{\Br}{{\rm Br}}
\newcommand{\Lag}{\mathcal{L}}
\newcommand{\Lqcd}{\Lambda_{\rm QCD}}

\newcommand{\ji}{j_{\rm inf}} 
\newcommand{\jb}{j_{B-L}} 
\newcommand{\M}{M} 
\newcommand{\im}{{\rm Im} }
\newcommand{\re}{{\rm Re} }

\def\lrf#1#2{ \left(\frac{#1}{#2}\right)}
\def\lrfp#1#2#3{ \left(\frac{#1}{#2} \right)^{#3}}
\def\lrp#1#2{\left( #1 \right)^{#2}}
\def\REF#1{Ref.~\cite{#1}}
\def\SEC#1{Sec.~\ref{#1}}
\def\FIG#1{Fig.~\ref{#1}}
\def\EQ#1{Eq.~(\ref{#1})}
\def\EQS#1{Eqs.~(\ref{#1})}
\def\TEV#1{10^{#1}{\rm\,TeV}}
\def\GEV#1{10^{#1}{\rm\,GeV}}
\def\MEV#1{10^{#1}{\rm\,MeV}}
\def\KEV#1{10^{#1}{\rm\,keV}}
\def\blue#1{\textcolor{blue}{#1}}
\def\red#1{\textcolor{blue}{#1}}

\newcommand{\eff}{\Delta N_{\rm eff}}
\newcommand{\neff}{\Delta N_{\rm eff}}
\newcommand{\cc}{\Omega_\Lambda}
\newcommand{\Mpc}{\ {\rm Mpc}}
\newcommand{\Msolar}{M_\odot}

\def\my#1{\textcolor{blue}{#1}}
\def\MY#1{\textcolor{blue}{[{\bf MY:} #1}]}

%%%%%%%%%%%%%%%%%%%%%%%%%%%%%%%%%%%%%%%%%%%%%%%%%%%%%%%%%%%%%%%

%######################
\begin{flushright}
TU-1266\\
KEK-QUP-2025-0014
\end{flushright}
%######################

\title{
PQ-ball and its Real Scalar Analogue in an Expanding Universe
}

\author{
Kazunori Nakayama$^{1,2}$,
Masaki Yamada$^{1}$}
\affiliation{$^{1}$Department of Physics, Tohoku University, Sendai, Miyagi 980-8578, Japan}
\affiliation{$^{2}$International Center for Quantum-field Measurement Systems for Studies of the Universe and Particles (QUP), KEK, Tsukuba, Ibaraki 305-0801, Japan}

\abstract{
We demonstrate the formation of quasi-stable localized scalar configurations in spontaneously symmetry breaking U(1) model by 3+1-dimensional classical lattice simulations. Such configurations are called PQ-balls, as the primary motivation of this kind of configuration is Peccei-Quinn theory under the kinetic misalignment mechanism. Our numerical simulations demonstrate that they can form if the PQ charge is generated through the coherent rotation of a complex scalar field in the complex plane, via dynamics analogous to the Affleck-Dine mechanism. These configurations subsequently decay due to the U(1)-breaking effect induced by spontaneous symmetry breaking. We also demonstrate the formation and decay of oscillons in a similar setup in a real scalar field theory. 
}

\emailAdd{kazunori.nakayama.d3@tohoku.ac.jp}
\emailAdd{m.yamada@tohoku.ac.jp}

\maketitle
\flushbottom

%%%%%%%%%%%%%%%%%%%%%%%%%%%%%%%%%%%%%%%%%%%%%%%%%%%
\section{Introduction}
%%%%%%%%%%%%%%%%%%%%%%%%%%%%%%%%%%%%%%%%%%%%%%%%%%%

Scalar fields play an important role in cosmology, as they can exhibit coherent motion during the evolution of the Universe.
In certain cases, scalar fields can support localized configurations that are energetically favored compared to a collection of free particle states.
This occurs in complex scalar field theories when the scalar potential is shallower than quadratic.
Such configurations are known as Q-balls~\cite{Coleman:1985ki}, whose stability is ensured by the conservation of a global U(1) charge.
In particular, Q-balls naturally form in the context of Affleck-Dine baryogenesis~\cite{Affleck:1984fy,Dine:1995kz}, where a baryon charge is generated through the coherent motion of a complex scalar field.
Perturbative analyses, as well as numerical simulations, have shown that small fluctuations grow and lead to Q-ball formation in nearly homogeneous, finite-charge systems~\cite{Kusenko:1997ad,Kusenko:1997zq,Kusenko:1997si,Enqvist:1997si,Enqvist:1998en,Kasuya:1999wu,Kasuya:2000wx,Kasuya:2000sc}.
Understanding their formation and subsequent evolution is important for cosmology, as Q-balls can alter the expansion history of the Universe and influence how and when particles thermalize with the ambient plasma.

Recently, it has been argued that quasi-stable, localized configurations can also exist even when the scalar potential has a tachyonic mass term at the origin, that is, when the vacuum exhibits spontaneous symmetry breaking (SSB)~\cite{Kobayashi:2025qao}%
\footnote{
See also Ref.~\cite{Kawasaki:2025nsi} for a related study in a two-field model. 
}
(see also Refs.~\cite{Nugaev:2016wyt,Nugaev:2019vru}).
Indeed, it is reasonable to expect the formation of Q-ball-like configurations if the SSB scale is much smaller than the amplitude of the coherent field motion.
We refer to such configurations as PQ-balls, since they are primarily motivated by the axion kinetic misalignment mechanism~\cite{Co:2019jts,Chang:2019tvx} for the QCD axion under the Peccei-Quinn (PQ) mechanism~\cite{Peccei:1977hh,Peccei:1977ur}, in which the U(1)$_{\rm PQ}$ charge is generated through dynamics analogous to those in Affleck-Dine baryogenesis.
In Ref.~\cite{Kobayashi:2025qao}, we discussed the formation and decay of PQ-balls under the assumption of spherical symmetry.
Such configurations can remain stable in a finite-density environment, while they gradually decay in vacuum.

In this paper, we demonstrate the formation and subsequent decay of PQ-balls in an expanding Universe using $3+1$-dimensional classical lattice simulations.
We specifically consider a complex scalar field with a logarithmically dependent mass term, motivated by the radiative generation of the PQ-breaking scale via renormalization group running.
Starting from a coherently rotating complex scalar field, we find that small perturbations grow and lead to PQ-ball formation.
At the time of formation, the average field amplitude in the outer region remains nonzero, meaning that PQ-balls are initially embedded in a finite-density background.
As the Universe expands, the average field amplitude redshifts and eventually falls below a critical threshold.
When this occurs, PQ-balls become unstable and begin to decay.
We show that the lifetime of a PQ-ball is longer for smaller values of the SSB scale.

As a complementary study, we also demonstrate the formation and decay of oscillons in a real scalar field theory, even when the potential exhibits spontaneous symmetry breaking at the vacuum.
Although oscillons are generally unstable, the SSB potential introduces a new decay channel that shortens their lifetime.

The remainder of this paper is organized as follows.
In Sec.\ref{sec:model}, we introduce the model and review the properties of PQ-balls discussed in our previous work.
In Sec.\ref{sec:PQball}, we present our numerical results on the formation and decay of PQ-balls.
Sec.~\ref{sec:oscillon} is devoted to the formation and decay of oscillons in a real scalar field theory.
Finally, in Sec.~\ref{sec:discussions}, we discuss the cosmological implications of our finidings.

\section{Model}
\label{sec:model}

We consider a Q-ball-like solution in a theory with spontaneous breaking of a global U(1) symmetry in vacuum, within an expanding spacetime.
In particular, we consider the following potential:
\begin{equation}
	V(\Phi) = m^{2}\left[ 1+k_{1}\log\frac{|\Phi|^{2}}{M^{2}} - k_{2}\log\frac{|\Phi|^{2}+v^{2}}{M^{2}} \right] |\Phi|^{2},
	\label{eq:potential_PQ}
\end{equation}
where $k_{2}>k_{1}>0$, 
arising from the beta function of the running mass. The parameter $M$ denotes the renormalization scale.
This potential leads to a spontaneously broken minimum located at (see, e.g., Ref.~\cite{Kobayashi:2025qao}) 
\beq
 f_a \simeq M \left( \frac{v}{M} \right)^{\frac{k_{2}}{k_{1}}} \exp\left(-\frac{(1+k_{1})}{2k_{1}}\right),
\eeq
which is significantly smaller than the other energy scales, such as $v$ and $M$, for small values of $k_1$. 
A relatively small value of SSB scale is realized from a large cutoff without small parameters or hierarchical dimensionfull parameters. 
Such a mechanism of “radiative stabilization” of the PQ field has been previously discussed in Refs.~\cite{Abe:2001cg,Nakamura:2008ey,Moroi:2014mqa}. 
Here, we extend the original potential so that the sign of the beta function for the running mass changes around the energy scale $v$.

\subsection{Evolution of averaged field value}

We are interested in the scenario where the scalar field initially acquires a large vacuum expectation value (VEV) and subsequently begins to rotate in field space at some point in time.
Although we do not delve into the specific mechanism responsible for this behavior, such dynamics can naturally arise in the context of the Affleck-Dine mechanism~\cite{Affleck:1984fy,Dine:1995kz} to generate a PQ charge in the axion kinetic misalignment mechanism~\cite{Co:2019jts,Chang:2019tvx}.

After the onset of rotation, the amplitude of the field gradually decreases due to cosmic expansion.
If the effect of the running mass in the scalar potential is neglected, the zero mode of the field amplitude evolves as $\left< \Phi(t) \right> \propto a^{-3/2}$
in an expanding Universe, where $a$ is the scale factor.
Since we are interested in the regime $k_1, k_2 \ll 1$, 
and the logarithmic correction varies only slowly, this scaling behavior is expected to remain approximately valid in our model as long as $\left< \Phi (t) \right> \gg f_a$.

In this paper, we assume that the initial amplitude of the scalar field is sufficiently larger than $v$. 
For sufficiently large $\abs{\Phi}$ compared to $v$, the potential simplifies to
\begin{equation}
	V(\Phi) \simeq m^{2}\left[ 1+(k_{1}-k_{2})\log\frac{|\Phi|^{2}}{M^{2}} \right] |\Phi|^{2},
\end{equation}
This form of the potential is known as the gravity-mediation type in the context of Affleck-Dine baryogenesis in supersymmetric theories.
In this case, it is known that a localized configuration, known as a Q-ball, exists when 
$k_{1}-k_{2}=K<0$. 
In fact, small density perturbations for the scalar field tends to grow after the onset of rotation for the scalar field, and they result in the formation of Q-balls. This has been confirmed by perturbative analysis as well as detailed numerical simulations~\cite{Kusenko:1997ad,Kusenko:1997zq,Kusenko:1997si,Enqvist:1997si,Enqvist:1998en,Kasuya:1999wu,Kasuya:2000wx,Kasuya:2000sc}. 

The above observation suggests that a quasi-stable, localized configuration can also exist in our original model, at least in the limit of small $v$. We refer to this configuration as a PQ-ball, 
where the U(1) symmetry is identified with the PQ symmetry.
Such a configuration remains stable in a finite-density plasma, while it gradually decays in vacuum. Its formation and decay have been discussed in Ref.~\cite{Kobayashi:2025qao}, under the assumption of spherical symmetry.

\subsection{Properties of PQ-ball}

We summarize some properties of the PQ-ball as discussed in Ref.~\cite{Kobayashi:2025qao}.

Even in a model with spontaneously broken U(1) symmetry, a stable localized configuration can exist in a finite-density environment.
For sufficiently small $v$, the configuration can be approximated by
\begin{equation}
 \Phi(r,t) \simeq \lmk \Phi_{0} \exp\left(-\frac{r^{2}}{R_{Q}^{2}}\right)  + \Phi_{\infty} \rmk e^{-i \omega t},
\end{equation}
where $\omega$ denotes the chemical potential and $R_\text{Q}$ is the Gaussian radius. 
These quantities are given by
\beq
 \omega^2 \sim \left(1+2(k_2-k_1)\right)m^2 
\\
	R_{Q} \simeq \frac{\sqrt{2}}{m\sqrt{k_{2}-k_{1}}}.
    \label{eq:gauss_rad}
\eeq
The critical value of the field amplitude, $\Phi_{\infty}$, is given by 
\beq
 \Phi_{\infty} \simeq M \lmk \frac{v}{M} \rmk^{\frac{k_{2}}{k_{1}}}\exp\left(\frac{1}{2k_{1}}(\omega^{2}/m^2-1-k_{1})\right).
	\label{eq:chi_inf}
\eeq
This corresponds to the asymptotic field amplitude at spatial infinity. Note that $\Phi_{\infty}$ is much larger than the spontaneous symmetry breaking scale $f_a$ in vacuum, 
and that the field configuration at $r \to \infty$ rotates in the complex plane as
$\Phi_{\infty} e^{-i \omega t}$. 
This indicates that the above solution represents a PQ-ball that resides in a finite-density environment.

The U(1) charge density is given by 
\begin{equation}
q(x) = i (\Phi^{*} \dot{\Phi}-\Phi \dot{\Phi}^{*}).
\label{Q_charge}
\end{equation}
Although the total U(1) charge is conserved, the charge contained within a PQ-ball can vary over time due to decay from its surface in vacuum.
For a spherically symmetric PQ-ball, we may define its U(1) charge as follows:
\begin{equation}
 Q = \int_0^{R_Q} 4\pi r^2 q(r) dr \,. 
\end{equation}
This can be approximated as 
\begin{equation}
	Q \sim \omega \Phi_0^2 R_Q^{3} .
\end{equation}

In an expanding Universe, the averaged field value $\left< \Phi (t)\right>$ decreases with time. 
Then it eventually falls below the threshold value $\Phi_{\infty}$. 
At that point, if PQ-balls are present, they become unstable and begin to decay.
According to Ref.~\cite{Kobayashi:2025qao}, the lifetime of the PQ-ball can be estimated by evaluating the outgoing charge flux at its surface, given by
\begin{equation}
	\frac{dQ}{dt} \sim -4\pi v_{q}\omega R_Q^{2}\Phi_{\infty}^{2},
	\label{eq:decayrate}
\end{equation}
where the parameter $v_{q}$ denotes the velocity of the outgoing charge flow and is expected to be of order unity.
Omitting $\mathcal{O}(1)$ numerical factors, we obtain 
\beq
	\left|\frac{1}{Q}\frac{dQ}{dt}\right| 
    &\sim& \frac{1}{R_Q}\left(\frac{\Phi_{\infty}}{\Phi_{0}}\right)^{2} 
    \\
    &\sim& m \sqrt{k_{2}-k_{1}} \left(\frac{M}{\Phi_{0}}\right)^{2} 
    \left( \frac{v}{M} \right)^{\frac{2k_{2}}{k_{1}}}\exp\left(\frac{1}{k_{1}}(2k_{2}-3k_{1})\right).
	\label{eq:rateratio}
\eeq
This expression gives the typical inverse timescale for the complete decay of a PQ-ball. 
Notably, this rate is proportional to
$(v/M)^{2 k_2/k_1}$ rather than $(v/M)^2$ or $(f_a/M)^2$.

It is worth noting that if $v$ is sufficiently small, the lifetime of a PQ-ball can be much longer than the timescale at which the field amplitude in the outer region reaches the critical value.
In such a case, the decay rate of the PQ-ball is governed by its intrinsic dynamics, rather than by the Hubble expansion.
This condition can be expressed as
\beq
 \abs{\frac{1}{Q} \frac{dQ}{dt}} \ll H(t_{\rm th}) 
\eeq
where $t_{\rm th}$ denotes the time when  $\left< \Phi (t) \right> = \Phi_{\infty}$. 
Our focus is on scenarios in which this condition is satisfied.

\section{Numerical simulations for PQ-ball formation} 
\label{sec:PQball}

The aim of this paper is to simulate the formation and decay of PQ-balls without assuming spherical symmetry, while taking into account cosmic expansion in an FLRW Universe.

The equation of motion in an expanding Universe is given by
\beq
	\ddot{\Phi} + 3 H \dot{\Phi}     - \nabla^2 \Phi + \frac{\partial V}{\partial \Phi^*} = 0 
\label{eq:EL2}
\eeq
where a dot denotes a derivative with respect to cosmic time, $H = \dot{a}/a$ is the Hubble parameter, and $a$ is the scale factor. 
Throughout this paper, we assume 
$a \propto t^{2/3}$ and $H = 2/(3t)$, which corresponds to a matter-dominated epoch (specifically, an inflaton-oscillation-dominated phase after inflation but before the completion of reheating).
This cosmological background is indeed realized in scenarios such as the Affleck-Dine mechanism to generate PQ-charge in the context of the axion kinetic misalignment mechanism.

The simulation begins at the onset of rotation in field space. We assume that the initial VEV of the field is $M$, and that its rotation velocity is characterized by $m$. Accordingly, the initial condition is imposed at
$t = t_0$ such as 
\beq
 &&\phi(t_0) = M + \delta \phi
 \\
 &&\dot{\phi}(t_0) = i m M + \delta \dot{\phi}, 
\eeq
where $\delta \phi$ and $\delta \dot{\phi}$ represent quantum fluctuations satisfying  
\beq
 &&\left< \delta \phi({\bm k}) \delta \phi({\bm k}') \right> = \frac{1}{2 \omega_k} (2\pi)^3 \delta^{(3)} ({\bm k} + {\bm k}') 
 \\
 &&\left< \dot{ \delta \phi}({\bm k}) \dot{\delta \phi}({\bm k}') \right> = \frac{\omega_k}{2} (2\pi)^3 \delta^{(3)} ({\bm k} + {\bm k}')  
\eeq
with $\omega_k = \sqrt{\abs{\bm k}^2 + m^2}$.

We numerically compute the time evolution of the field.
To simplify the equations, dimensionful parameters are rescaled as
$\Phi \to M \Phi$, $v \to M v$, $t \to t/m$, and $x \to x /m$. 
We use a time step of
$\Delta t=0.01$ 
and simulate the system in a three-dimensional box of volume $L^3=20^3$, employing a sixth-order symplectic integration method.
The spatial grid spacing is
$\Delta x = L / N$, with $N^3 = 256^3$ 
grid points in total.
Periodic boundary conditions are imposed at the boundaries of the simulation box.

As a benchmark case, we take
$k_{1}=0.1$ and $k_{2}=0.2$. 
The initial condition is set at $t = t_0 = 100$ with quantum fluctuations seeded using a method similar to that described in Ref.~\cite{Felder:2000hq} (see also Refs.~\cite{Kawasaki:2011vv,Hiramatsu:2013qaa,Kamada:2014qja,Kamada:2015iga}). 
We introduce a cutoff to suppress high-frequency modes with
$\abs{{\bm k}} \gtrsim m$. 
After the rescaling, the ratio
$m/M$ 
appears only in the amplitude of the quantum fluctuations.
In this work, we adopt
$m/M = 0.1$ 
as an example.
We note that the precise form of the initial fluctuations does not qualitatively affect our results.
The timescale for PQ-ball formation depends only logarithmically on the initial fluctuation amplitude, and the decay process of the PQ-ball is insensitive to the initial conditions once the PQ-balls have formed.

\begin{figure}[t]
	\centering	\includegraphics[width=0.4\linewidth]{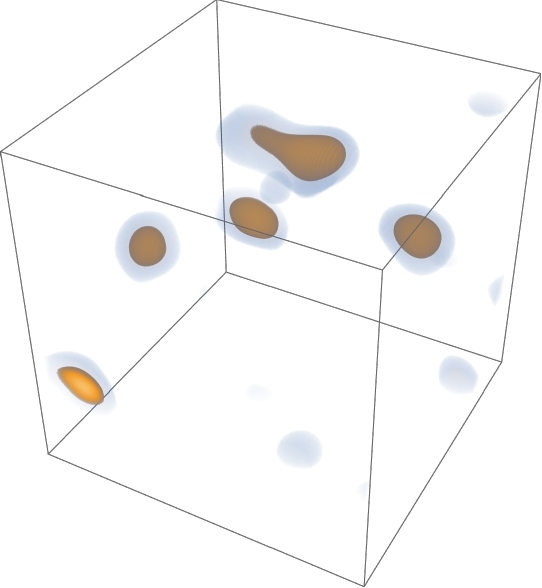}
    \hspace{0.3cm}
     \includegraphics[width=0.4\linewidth]{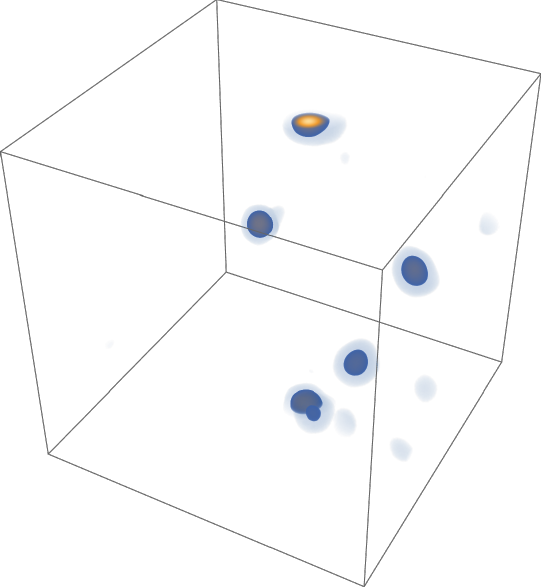}
     \vspace{0.5cm}\\
     \includegraphics[width=0.4\linewidth]{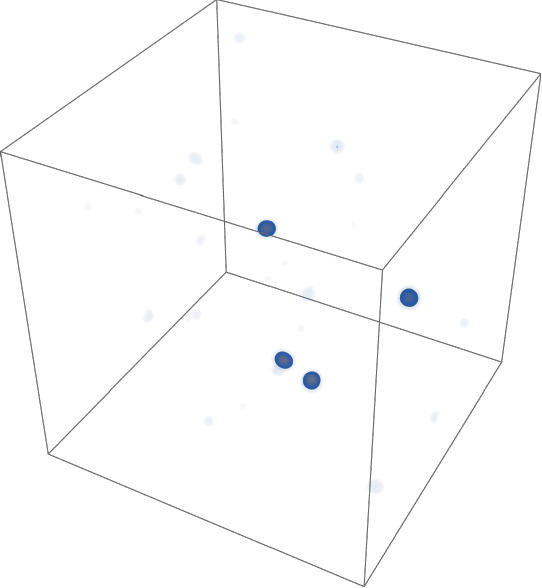}
    \hspace{0.3cm}
     \includegraphics[width=0.4\linewidth]{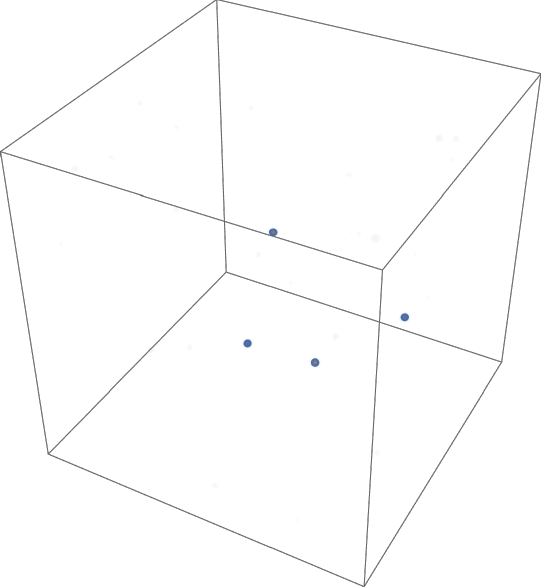}
	\caption{
    Snapshots for the U(1) charge density distribution at $t= 300$ (top left), $1000$ (top right), $3000$ (bottom left), $10000$ (bottom right) for the case with $v = 0$. 
    }
	\label{fig:evolution1}
\end{figure}

\begin{figure}[t]
	\centering	\includegraphics[width=0.4\linewidth]{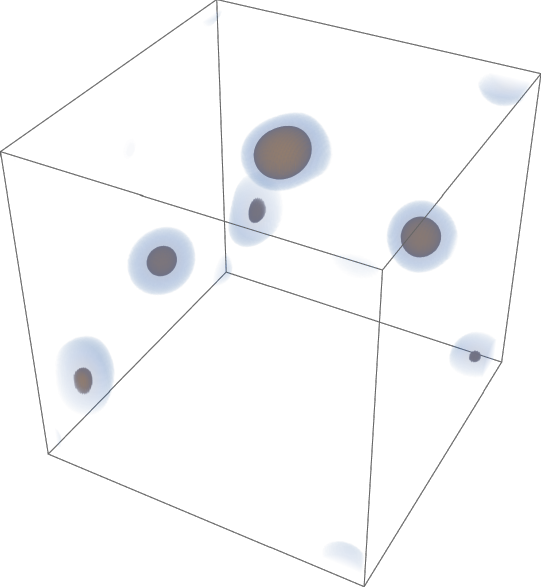}
    \hspace{0.3cm}
     \includegraphics[width=0.4\linewidth]{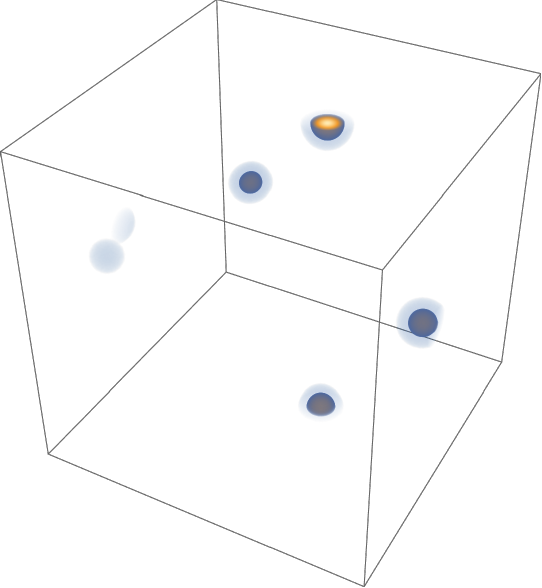}
     \vspace{0.5cm}\\
     \includegraphics[width=0.4\linewidth]{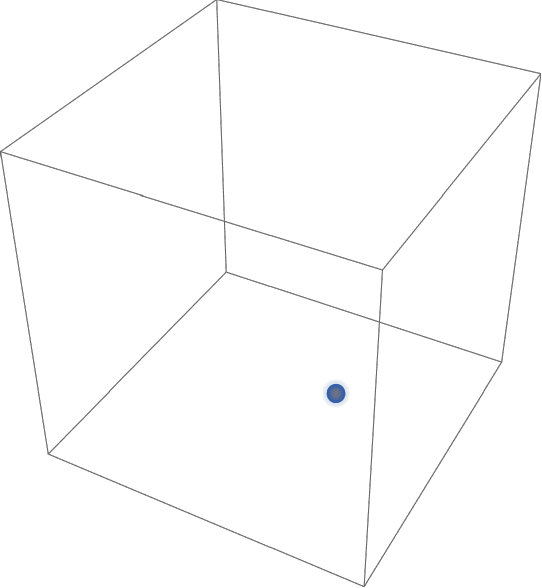}
    \hspace{0.3cm}
     \includegraphics[width=0.4\linewidth]{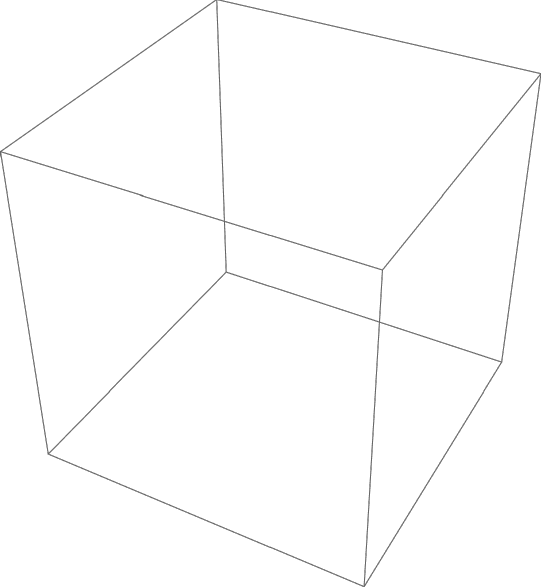}
	\caption{
    Same with Fig.~\ref{fig:evolution1} but with $v = 0.3$. 
    }
	\label{fig:evolution2}
\end{figure}

We identify PQ-balls in our numerical simulations as regions where the U(1) charge density exceeds ten times the average U(1) charge density. 
Figures~\ref{fig:evolution1} and \ref{fig:evolution2} present snapshots of the PQ-ball at $t= 300$ (top left), $1000$ (top right), $3000$ (bottom left), $10000$ (bottom right) for the cases with $v = 0$ and $0.3$, respectively. 
The shaded regions indicate areas where the U(1) charge density exceeds ten times the average value.%
\footnote{
Note that the shaded region does not represent the entire PQ-ball, but rather only its core.
This is because we plot only those regions where the U(1) charge density exceeds ten times the average value in the outer region.
Even if the dense region appears as a small point at $t=10000$ in Fig.~\ref{fig:evolution1}, the resolution of our numerical simulation is sufficiently high to resolve the internal structure of the PQ-ball. 
}
At early times (e.g., $t = \mathcal{O}(100)$), the charge distributions are similar for both cases.%
\footnote{
Note that the same initial condition is used for different values of $v$. 
}
At later times, however, only relatively large PQ-balls survive in the
$v = 0.3$ case, and they eventually disappear by $t \sim 4000$.
It is important to note that the simulation box has a fixed comoving size; thus, PQ-balls with a fixed physical size will appear smaller as time progresses.
As we will explicitly show below, PQ-balls do not decay in the case of $v=0$.
These results demonstrate the formation and decay of PQ-balls in a $3+1$-dimensional lattice simulation, where stable Q-balls form in the limit $v \to 0$.

\begin{figure}[t]
	\centering	\includegraphics[width=0.6\linewidth]{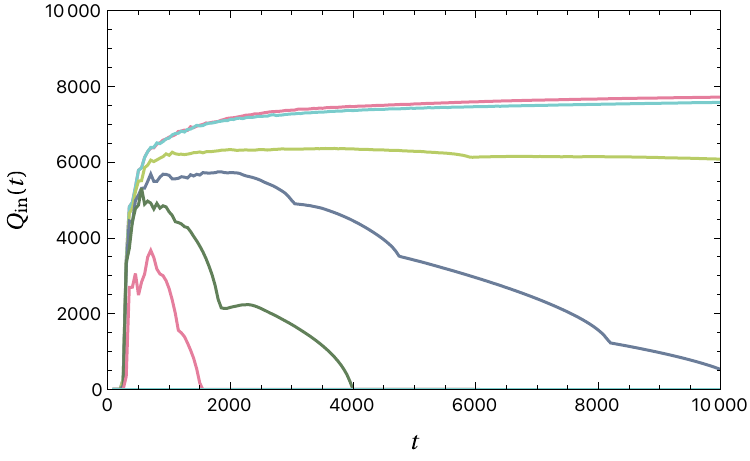}
	\caption{
    Time evolution of total U(1) charge inside PQ-balls for the case with $v = 0$, $0.01$, $0.1$, $0.2$, $0.3$, $0.4$ from top to bottom. 
    }
	\label{fig:PQball}
\end{figure}

Figure~\ref{fig:PQball} shows the time evolution of the total charge contained within the PQ-balls, which we denote $Q_{\rm in}(t)$.
It demonstrates that PQ-balls form around 
$t = \mathcal{O}(100)$ and remain stable for sufficiently small $v$. 
For $v \gtrsim 0.1$, 
however, the energy density of PQ-balls decreases over time as they decay into the surrounding region during the timescale of the simulation.
During the decay phase, the charge
$Q_{\rm in}(t)$ exhibits several non-monotonic features—e.g., a noticeable bump around $mt \sim 2000$ for $v = 0.3$.
To understand these features, it is important to note that PQ-balls with various sizes form in the simulation.
Smaller PQ-balls tend to decay earlier, releasing their charge into the surrounding environment.
Due to the finite box size of the simulation, the decay products can subsequently be absorbed by larger, surviving PQ-balls, leading to temporary increases in $Q_{\rm in}(t)$. 

Eventually, all PQ-balls fully decay at approximately
$t = 10^4$ for $v = 0.2$, $t=4000$ for $v = 0.3$, and $t=1500$ for $v = 0.4$. 
The lifetimes of the PQ-balls are consistent with the analytical estimate given in Ref.~\cite{Kobayashi:2025qao}. 
In particular, the lifetime scales as $\propto v^{2k_2/k_1} = v^{4}$ from Eq.\eqref{eq:rateratio}, which is marginally consistent with our numerical results.

\section{Case with a real scalar theory: oscillon formation}
\label{sec:oscillon}

It is known that a quasi-stable, localized configuration, called an oscillon, can exist in a real scalar field theory if the potential has a structure similar to that of a complex scalar theory admitting Q-ball solutions~\cite{Bogolyubsky:1976nx,Segur:1987mg,Gleiser:1993pt,Copeland:1995fq,Kasuya:2002zs}.
Although oscillons are not exactly stable due to the absence of U(1) symmetry, they can be long-lived owing to the approximate conservation of an adiabatic invariant in the non-relativistic limit~\cite{Gleiser:2008ty,Saffin:2014yka,Mukaida:2014oza,Mukaida:2016hwd,Ibe:2019vyo,Zhang:2020bec}.
We therefore expect that a similar argument applies even in the case of a real scalar field theory with a weakly spontaneously broken potential.

In this section, we consider a real scalar field theory with a similar potential:
\begin{equation}
	V(\phi) = \frac{1}{2} m^{2}\left[ 1+k_{1}\log\frac{\phi^{2}}{M^{2}} - k_{2}\log\frac{\phi^{2}+v^{2}}{M^{2}} \right] \phi^{2},
	\label{eq:potential_PQ}
\end{equation}
In the case of $v=0$, if $\phi$ begins oscillating with a large amplitude,
small perturbations can grow and lead to the formation of quasi-stable localized configurations known as oscillons.
Oscillons are generally not absolutely stable, but in the presence of a gravity-mediation-type potential as above, they can be extremely long-lived.

Similar to the complex scalar field case that gives rise to PQ-ball formation,
we expect that oscillons also form when 
$v$ is small but nonzero.
We further expect that these oscillons decay due to the symmetry-breaking term introduced by the nonzero
$v$, 
even though oscillons themselves are not strictly stable even in the $v=0$ case.
In other words, the lifetime of oscillons becomes shorter for $v \ne 0$, and their decay behavior is expected to follow a relation similar to \eq{eq:decayrate}.

\begin{figure}[t]
	\centering	\includegraphics[width=0.4\linewidth]{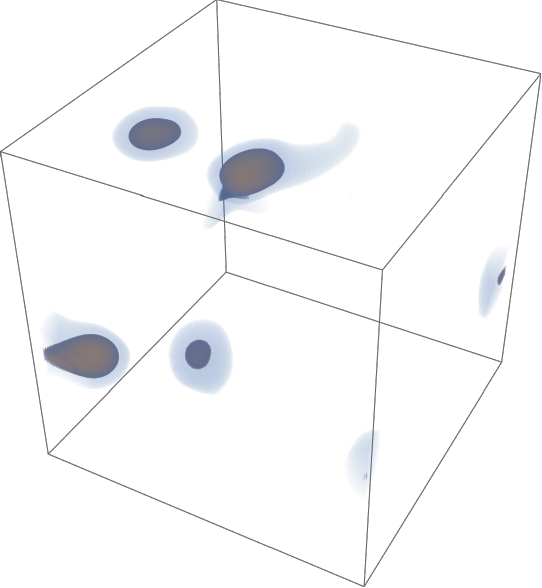}
    \hspace{0.3cm}
     \includegraphics[width=0.4\linewidth]{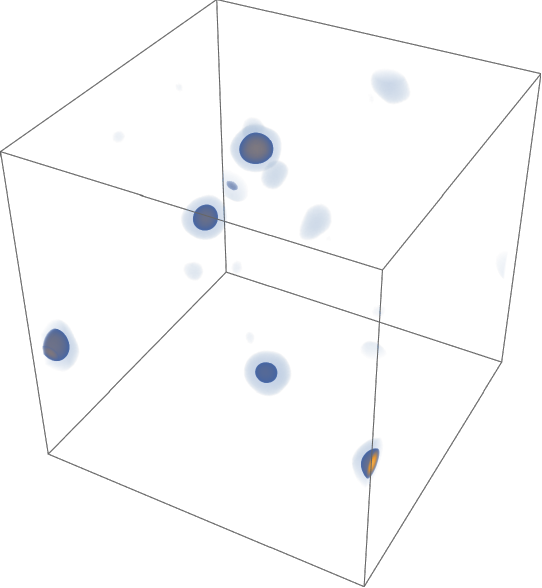}
     \vspace{0.5cm}\\
     \includegraphics[width=0.4\linewidth]{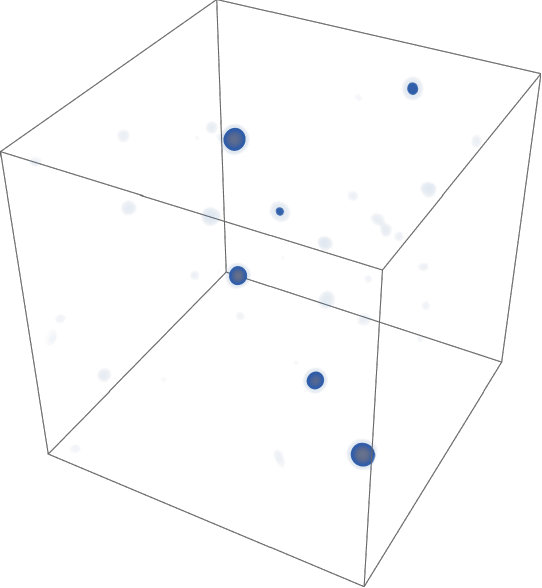}
    \hspace{0.3cm}
     \includegraphics[width=0.4\linewidth]{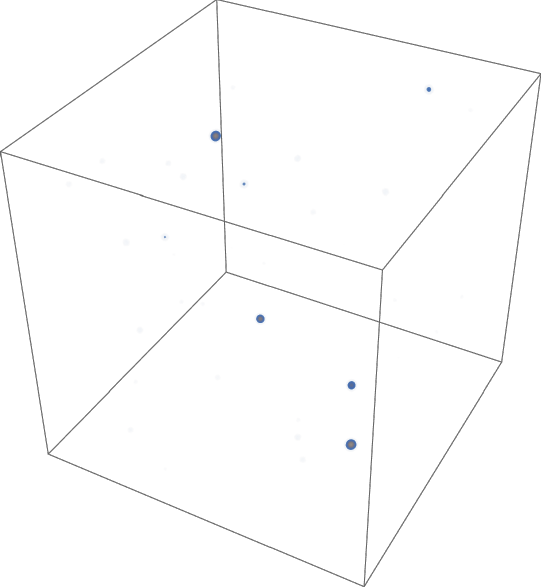}
	\caption{
    Snapshots for the energy density distribution in the real scalar field theory at $t= 300$ (top left), $1000$ (top right), $3000$ (bottom left), $10000$ (bottom right) for the case with $v = 0$. 
    }
	\label{fig:evolution3}
\end{figure}

\begin{figure}[t]
	\centering	\includegraphics[width=0.4\linewidth]{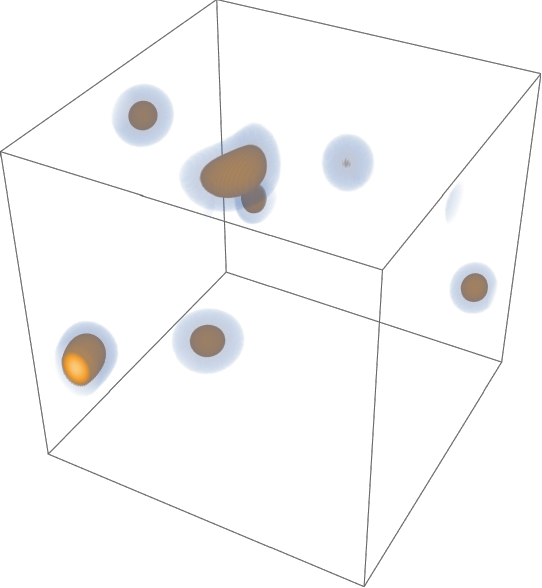}
    \hspace{0.3cm}
     \includegraphics[width=0.4\linewidth]{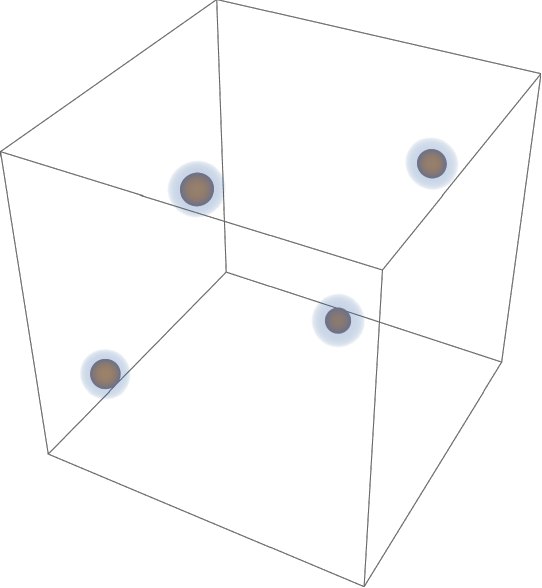}
     \vspace{0.5cm}\\
     \includegraphics[width=0.4\linewidth]{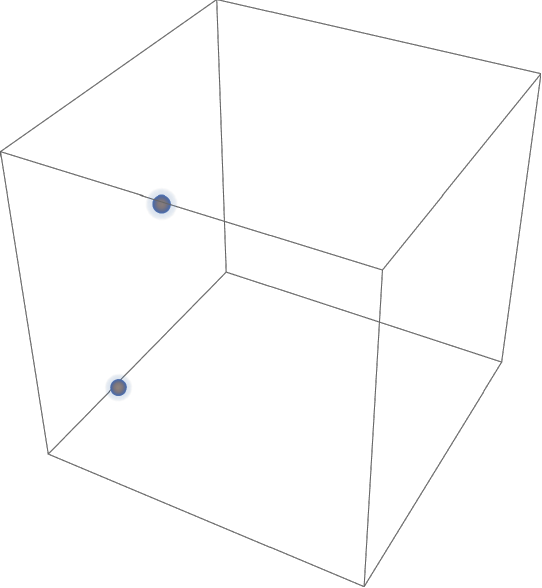}
    \hspace{0.3cm}
     \includegraphics[width=0.4\linewidth]{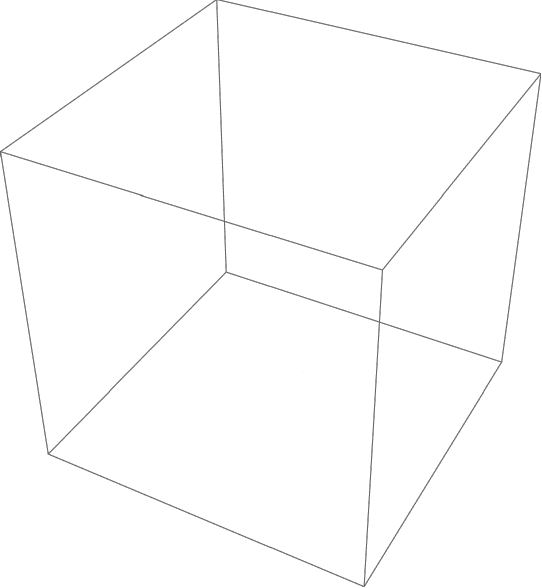}
	\caption{
    Same with Fig.~\ref{fig:evolution3} but with $v = 0.2$. 
    }
	\label{fig:evolution4}
\end{figure}

In our numerical simulations, we identify oscillons as regions where the energy density exceeds ten times the average energy density.
Figures~\ref{fig:evolution3} and \ref{fig:evolution4} show snapshots of the spatial energy density distribution at
$t= 300$ (top left), $1000$ (top right), $3000$ (bottom left), $10000$ (bottom right) for the cases $v = 0$ and $0.2$, respectively. The results are qualitatively similar to those in the complex scalar field theory: localized configurations form in both cases, but they decay by $t = 10000$ for the case of $v = 0.2$.

\begin{figure}[t]
	\centering	\includegraphics[width=0.6\linewidth]{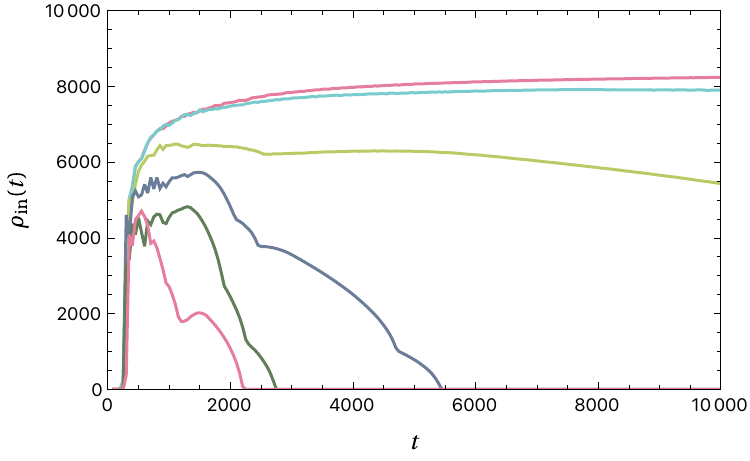}
	\caption{
    Time evolution of total energy density inside oscillons for the case with $v = 0$, $0.01$, $0.1$, $0.2$, $0.3$, from top to bottom. 
    }
	\label{fig:oscillon}
\end{figure}

Figure~\ref{fig:oscillon} shows the time evolution of the total energy within the oscillons, which we denote $\rho_{\rm in}(t)$.
Oscillons form around 
$t = \mathcal{O}(100)$ and exhibit long lifetimes when $v$ is small.
The overall behavior closely resembles that observed in the PQ-ball case.
We therefore conclude that similar localized objects—analogous to PQ-balls—form and decay even in a real scalar field theory.

\section{Discussions and conclusion}
\label{sec:discussions}

We have demonstrated the formation and decay of PQ-balls, quasi-stable, localized configurations arising in a complex scalar field theory with a spontaneously breaking potential.
Assuming that the scale of SSB is sufficiently small, PQ-balls can be long-lived and persist over cosmological timescales.
Indeed, by choosing the SSB scale appropriately, a single numerical simulation can capture both the formation and subsequent decay of PQ-balls from a coherently rotating scalar field configuration with small initial perturbations.
Although the PQ-balls formed have similar characteristic sizes, relatively small PQ-balls decay earlier, while larger ones survive for longer periods.
The time at which all PQ-balls disappear depends monotonically on the SSB scale, and this parameter dependence is qualitatively consistent with analytical expectations under the assumption of spherical symmetry discussed in our previous work.

We have also studied the formation and decay of oscillons in a real scalar field theory, where the potential can be regarded as a projection of the complex scalar potential onto the real axis.
Although oscillons are generally unstable, the presence of an SSB term introduces an additional decay channel, leading to shorter lifetimes for larger values of the SSB scale.
Note that this is not merely a case of oscillon formation in a 
$Z_2$-breaking potential, as often studied in the literature, where the scalar field oscillates around a global minimum with an amplitude smaller than or comparable to the SSB scale.
In our case, the amplitude of the scalar field configuration at the center of the oscillon is significantly larger than the SSB scale.

Our result of oscillon formation is nontrivial: in the complex scalar case, the field can rotate in the complex plane with a nearly constant amplitude and may never traverse the SSB potential barrier.
In contrast, the real scalar field necessarily crosses the SSB potential during each oscillation cycle.
Thus, there is no a priori reason to expect the effect of the SSB potential to be similar in both cases, yet our results show that the qualitative behavior is indeed similar.

Our findings indicate that quasi-stable, localized configurations tend to form in a broad class of models, both in complex and real scalar field theories.
This may have important implications for cosmology, particularly in the context of the axion kinetic misalignment mechanism for the QCD axion, where the thermalization timescale of the PQ-breaking field plays a critical role in determining the viability of the scenario.

\section*{Acknowledgements}
This work was supported by JSPS KAKENHI Grant Numbers 24K07010 [KN], 23K13092 [MY].
This work was supported by World Premier International Research Center Initiative (WPI), MEXT, Japan.

\bibliography{reference}

\end{document}